\documentclass[10pt,journal,twocolumn]{IEEEtran}

\usepackage[pdftex]{graphicx}
\graphicspath{{../}}
\DeclareGraphicsExtensions{.eps,.pdf,.jpeg,.png}
\usepackage{fixltx2e}
\usepackage{balance}
\usepackage{soul}

\usepackage{hyperref,color,soul,multirow}
\usepackage{url}
\usepackage{epstopdf}

\usepackage[cmex10]{amsmath}

\usepackage{float,multirow}
\usepackage{amsfonts}
\usepackage{comment}

\usepackage{epsfig,bm,amssymb,graphicx,setspace}
\usepackage{algorithmic,algorithm}
\usepackage{color}

\begin{document}

\title{Industrial Internet of Things (IIoT) Applications of Edge and Fog Computing: A Review and Future Directions}

\author{\IEEEauthorblockN{G. S. S. Chalapathi$^1$, Vinay Chamola$^2$, Aabhaas Vaish$^2$ and Rajkumar Buyya$^1$\\
		$^1$School of Computing and Information Systems, The University of Melbourne, Australia \\$^2$Department of Electrical and Electronics Engineering, BITS Pilani, Pilani Campus, India\\
}}

\maketitle

\begin{abstract}
With rapid technological advancements within the domain of Internet of Things (IoT), strong trends have emerged which indicate a rapid growth in the number of smart devices connected to IoT networks and this growth cannot be supported by traditional cloud computing platforms. In response to the increased capacity of data being transferred over networks, the edge and fog computing paradigms have emerged as extremely viable frameworks that shift computational and storage resources towards the edge of the network, thereby migrating processing power from centralized cloud servers to distributed LAN resources and powerful embedded devices within the network. These computing paradigms, therefore, have the potential to support massive IoT networks of the future and have also fueled the advancement of IoT systems within industrial settings, leading to the creation of the Industrial Internet of Things (IIoT) technology that is revolutionizing industrial processes in a variety of domains. In this paper, we elaborate on the impact of edge and fog computing paradigms on IIoT. We also highlight the how edge and fog computing are poised to bring about a turnaround in several industrial applications through a use-case approach.  
Finally, we conclude with the current issues and challenges faced by these paradigms in IIoT and suggest some research directions that should be followed to solve these problems and accelerate the adaptation of edge and fog computing in IIoT. 
\end{abstract}

\IEEEpeerreviewmaketitle

\section{Introduction}
\label{Intro}
\textbf{The Internet of Things (IoT)} \cite{Buyya-gubbi} refers to a system of smart devices which are connected to each other through the Internet. The basic structure of IoT systems involves the use of a large number of smart devices which are able to acquire, process, transmit and receive data between one another thereby enabling us to reliably monitor and precisely control any environment, control system or device through this system of interconnected smart devices. With forecasts predicting an estimated 28.5 billion network-connected devices to become active by 2022 \cite{devices0}, the IoT technology is poised to make a total economic impact between \$3.9 trillion and \$11.1 trillion per year in 2025 \cite{econ0}. While most of the IoT systems developed until now have been consumer-centric, the disruptive nature of this technology has enabled the adoption of this technology in a gamut of industrial settings thus leading to the development of \textbf{Industrial Internet of Things (IIoT) technology} \cite{intro0}. IIoT technology, in essence, refers to a system of interconnected smart devices in an industrial setting which connects industrial resources including sensors, actuators, controllers, machines with each other as well as with intelligent control systems which analyze the acquired data and optimize the ongoing industrial processes in order to improve execution speed, reduce involved costs, and dynamically control the industrial environment \cite{intro0}.

{One of the most important reasons behind the meteoric rise of IIoT systems in various industries is that IIoT systems can lead to a significant improvement in efficiency, throughput, and response time of operations inside these industries \cite{industries0}. IIoT has already revolutionized companies in many major industries across the globe, including the mining industry where IIoT systems have led to the installation of wireless access points in mining tunnels and RFID tracking technology has helped companies in tracking vehicles leading to an increase in production levels by 400\% \cite{cis0}. Proposed IIoT systems in agricultural settings can help farmers in nutrient monitoring as well as automated irrigation to improve crop yield \cite{use1}. The medical industry can also benefit from the capabilities of Industrial IoT systems where emergency services can access data from patients, ambulances and doctors to help all stakeholders in making informed decisions and improve resource utilization \cite{use2}. Pilot projects in China have successfully implemented an NB-IoT (Narrow Band IoT) system for smart electrical meters which allows real-time collection of power consumption data thereby enabling the energy grid officials to improve the electricity supply strategy in any area \cite{use3}. Similarly, NB-IoT smart parking systems have been deployed in cities to help drivers easily find parking spaces while integration of this system with payment solutions has led to automated transaction authorization for parking payment which has subsequently improved utilization of parking bays \cite{use4}. The railway industry can also leverage the power of IIoT solutions to improve the functioning of surveillance systems, signalling systems, predictive maintenance and Passenger or Freight Information Systems in order to improve services and safety \cite{use5}. Supply Chain Management (SCM) can also benefit by adopting IIoT based systems which will directly enhance tracking and traceability while also aiding in the optimization of shipment routes based on rapidly changing customer requirements \cite{use6}.}  

\begin{figure*}[ht]
	\includegraphics[width=\linewidth]{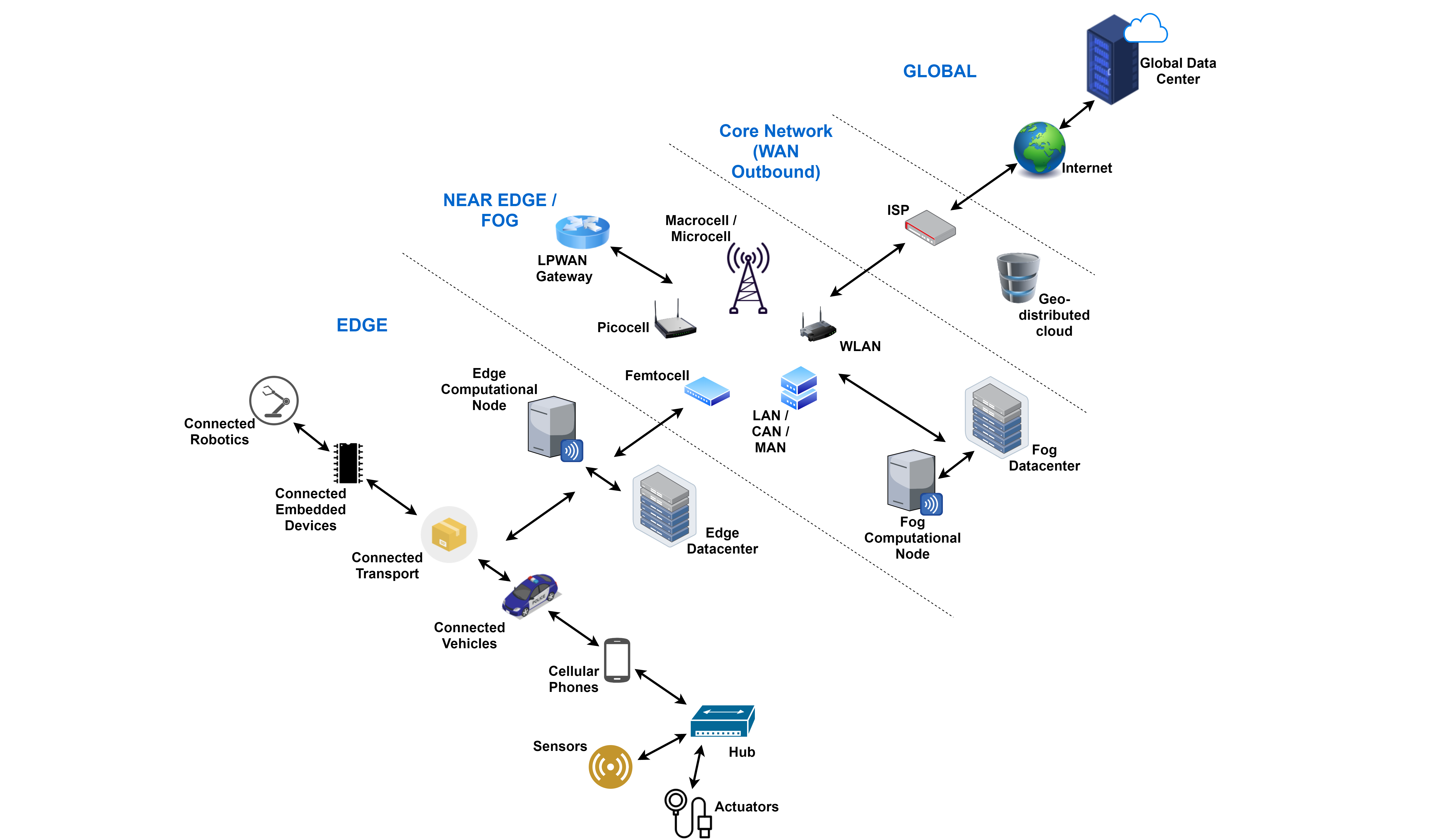}
	\caption{Edge, Fog, and Cloud Tiers} 
	\label{fig_2t}
	\vspace{-0.1in}
\end{figure*}
While the IIoT shows immense potential as a transformative technology, it is important to know the critical requirements that must be validated and verified in the design of IIoT systems so as to maximize the efficiency and performance of these systems \cite{req1,req2}. These requirements arise from the challenges often faced by Cyber Physical Systems (CPS) and they include the following:
\begin{enumerate}
\item {Scalability}
\item {Fault Tolerance or Reliability}
\item {Data Security}
\item {Service Security}
\item {Functional Security}
\item {Data Production and Consumption Proximity}
\end{enumerate}

With the rise in computational power being offered by systems in recent years, the focus of most industries has shifted towards garnering practical and useful patterns from their data which has been aided by the rapid development in statistical analysis and learning-based algorithms. Today, industries that are making use of IIoT solutions want to utilize the massive amount of data being generated to collect useful insights which can help in reduction of unplanned down-times, improve efficiency of production, lower energy consumption, etc. However, in order to process such massive amounts of data, IIoT systems generally require cloud computing services which often experience large round-trip delays and poor Quality of Service (QoS) as a large amount of data needs to be transferred to centralized data-centres for computation \cite{latency0}. Since most sensors and data acquisition devices in IIoT systems operate at the periphery of the network, more data tends to be produced near the periphery of the network, which implies that processing the data at the edge of the network would be more efficient \cite{edgeovercloud}. Therefore, efforts in shifting the computational power towards the periphery of the network have given rise to the edge and fog computing paradigms.

\textbf{Edge Computing} refers to the computing paradigm in which computations are performed at the edge of the network instead of the core of the network. In this scenario, the "edge" refers to any resource located on any network path between data acquisition devices (situated near the periphery of the network) and the cloud data-centre (situated at the core of the network) \cite{edgeovercloud}. The basis of the edge computing paradigm is that the computations should be done on the "edge" which is in proximity of the data sources and this avoids the latency associated with data transfer to the network's core.

The \textbf{Fog Computing} paradigm is similar in nature to edge computing in that it also has a decentralized architecture for computation but with the fundamental difference being that Fog Computing can expanded to the core of the network as well \cite{fog0}. This means that resources located at both edge and core can be used for computations and consequently, fog computing can aid in the development of multi-tier solutions which can offload service demand to the core of the network as the load \cite{fog0}. However, in most fog computing systems, the computational power is concentrated with the LAN resources which are closer to the data sources and further away from the network core, thus reducing the latency associated with data transfer to the core as seen in edge computing as well. Therefore, the fundamental difference between the edge and fog computing paradigms is basically in the location where the computational power and intelligence is stored. In case of edge computing, this computational power is concentrated at the edge of the network usually in powerful embedded devices like wireless access points or bridges whereas in the case of fog computing, the compute power is usually in the LAN resources. 
The rest of the paper is organized as follows: Section II discusses the background of edge and fog computing systems and how these paradigms address the requirements of modern IIoT systems. Section III describes various applications of edge computing in industrial settings. Section IV elaborates on fog computing applications. In Section V we present several outstanding issues and challenges with these computing paradigms that can be interpreted as future directions for research in this domain. Finally, in Section VI we conclude with the salient points of this paper.

\begin{figure*}[ht]
	
	\includegraphics[width=\linewidth]{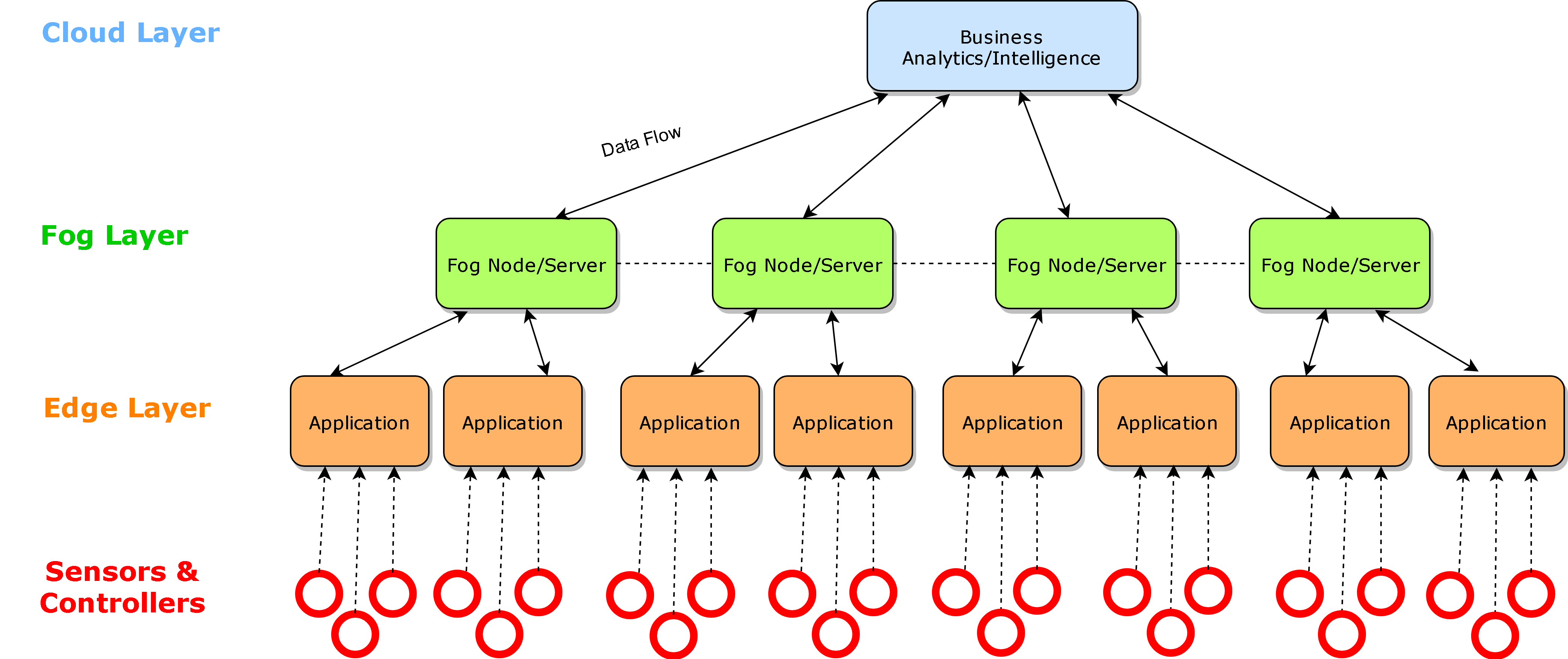}
	\caption{Industrial IoT Data Processing Layer Stack} 
	\label{fig_1t}
	\vspace{-0.1in}
\end{figure*}

\section{Relevant computing Paradigms and Requirements}
{The edge computing paradigm is a computing technology which enables data to be processed almost exclusively on the "edge" of the network, which refers to locations between the end devices (like sensors, controllers, and actuators) and the centralized cloud servers. The rationale behind the development of this technology is that computations performed closer to the end devices will lead to a lower latency in the system. This is because the system does not need to transfer data between edge devices and central cloud servers as the computations have been offset to closer locations on the edge. Therefore, in edge computing systems, edge devices can not only request content and services from the cloud servers but can also perform computational offloading, caching, storage, and processing, thereby making the edge devices both data producers and consumers \cite{edgeovercloud}.}

{The fog computing paradigm can be understood as an extension of the traditional cloud computing model wherein additional computational, data handling, and networking resources (nodes) are placed at locations on the network which are in close proximity to the end devices \cite{fogtax0}. The consequence of this extension is that processes involving data management, data processing, networking, and storage can occur not only on the centralized cloud servers, but also on the connections between end devices and the cloud servers \cite{fogtax1}. Fog computing, therefore, can be extremely useful for low latency applications as well as applications that generate an enormous amount of data that cannot be practically transferred to cloud servers in real-time due to bandwidth constraints \cite{fogtax2}.}

{As discussed in the previous section, there are many requirements which cyber physical systems need to maintain so as to become a viable supplement for real-world operations and applications. These include the following:}
\begin{enumerate}

\item {{\textbf{Scalability} which ensures that the increased data transfer between nodes does not degrade latency or response time.}}
\item {{\textbf{Fault tolerance and reliability} which guarantees that the system functions normally under variable external factors like under high load conditions.}}
\item {{\textbf{Data security} which ensures that the system is resistant to external attacks attempting to steal confidential information stored in the system or network.}}
\item {\textbf{Service security} to make the system resistant to external attacks which are attempting at disrupting the service provided by the system to the industry such as through Denial-of-service (DoS) attacks or Blackhole attacks.}
\item {\textbf{Functional security} so that physical accidents such as fires, explosions, leaks do not occur at any time especially in industries handling potentially hazardous substances such as nuclear plants, chemical plants, and oil rigs}
\item {\textbf{Data production and computation proximity} which ensures that the devices collecting the data and the systems processing the data are close to each other over the network to reduce latency.}

\end{enumerate}
In order to realize the strengths offered by the edge and fog computing paradigms, IIoT systems must be designed in accordance with network structures of these paradigms since these paradigms adhere to all the requirements of cyber physical systems:

\begin{enumerate}
\item Edge and fog computing based systems are \textbf{scalable} since increased data transfer between nodes can be addressed by the introduction of additional edge devices to compensate for the added computational load without degrading the network's latency since these devices function in proximity to end devices, and hence, do not increase the data transfer delays over the network.

\item {Edge and fog computing systems are \textbf{reliable and fault tolerant} especially when compared with cloud-based systems since faults in the centralized cloud servers would result in a total loss of service but the decentralized nature of Edge and Fog Computing systems ensures that even if some of the computational nodes fail, the remaining healthy nodes can still maintain partial service. Furthermore, if the computational load of the failed nodes can be offset to the remaining healthy nodes, then the system can still run full service while corrective action is undertaken.}

\item {Edge and fog computing systems maintain \textbf{data security} within the system due to data decentralization which means that if an adversary wants to breach the system, it would need to breach each one of the large number of decentralized computing nodes in order to collect the entire system's data.}

\item {Edge and fog computing systems maintain \textbf{service security} by using advanced defense mechanisms such as per-packet-based detection, data perturbation, and isolation networks for the identification of and defense against attacks \cite{defense}.}

\item {Edge and fog computing systems ensure \textbf{functional security} since these systems as they can be used to create extremely stable and robust multi-loop control systems for functionally sensitive industrial operations such as temperature control \cite{tempcon}.}

\item {Edge and fog computing systems were developed with the rationale that \textbf{data consumption} (processing, storing, caching, etc.) and \textbf{production} are always in \textbf{proximity} which is ensured by the fundamental structure of these systems where computational nodes are located on the edges of the network, which are in close proximity to the end devices at the periphery of the network.}
\end{enumerate}
The distributed nature of edge and fog computing systems leads to several advantages in terms of reduced communication times and improved reliability, which makes these systems especially useful in a variety of industrial settings that require reliable, latency-sensitive networks for process automation. By realizing the inherent advantages of these paradigms, a large number of industries have started to utilize these paradigms in their system designs and we shall look at several such use cases in the following sections of this paper.

\section{Industrial Applications of Edge Computing}
\subsection{Manufacturing Industry}
{In order to understand the applications of edge computing in manufacturing, we will be considering the system architecture for a manufacturing-based setup as presented in Fig. \ref{fig_2}. After describing this architecture, a case study is presented which is based on the implementation of an active maintenance system on a prototype platform. Finally, this subsection concludes with a summary of the tests and results from this case study, as presented in \cite{manu}.}
\subsubsection{{System Architecture}}
\begin{figure*}[ht]
	\centering
	\includegraphics[width=160mm]{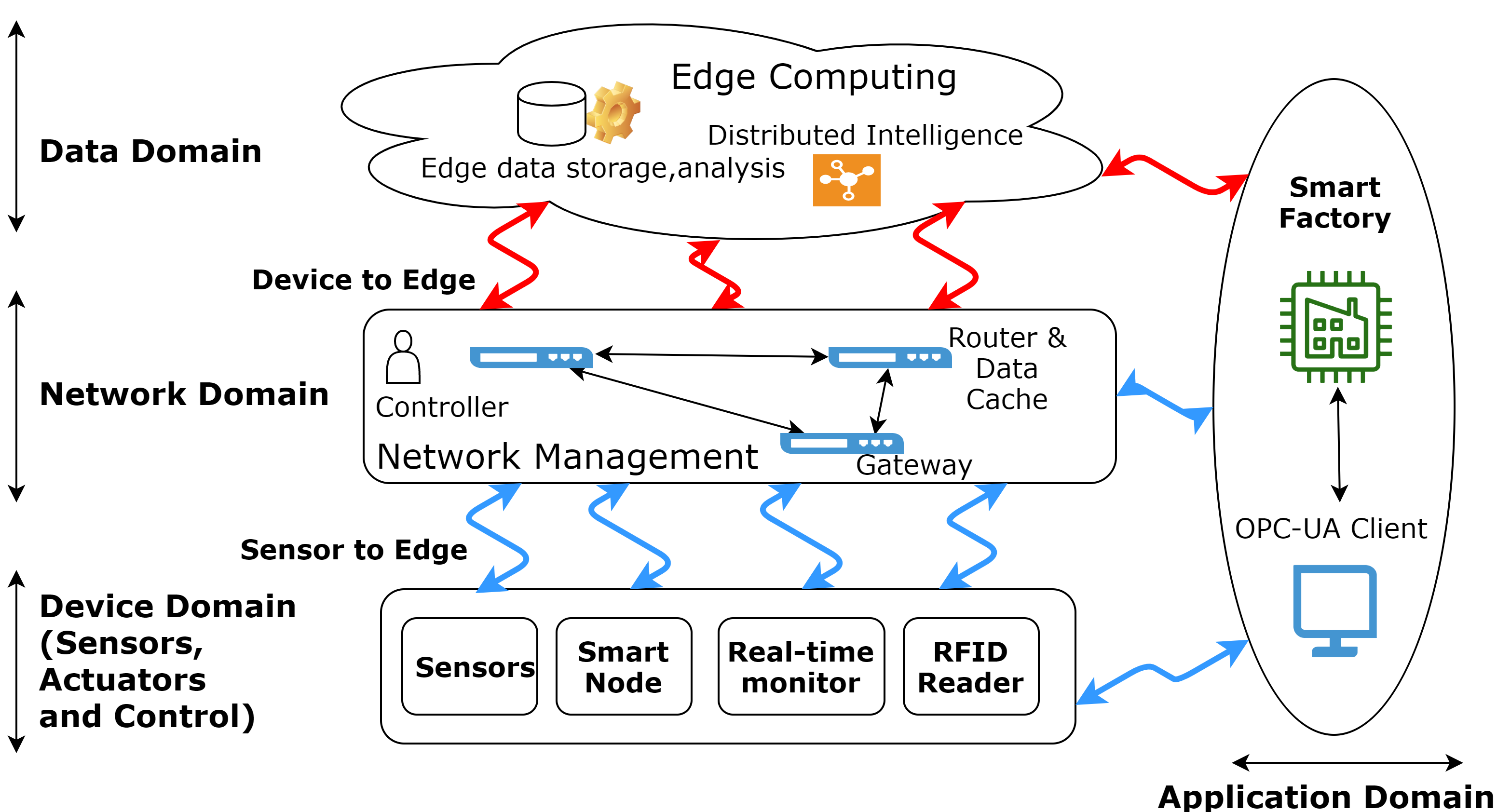}
	\caption{Architecture of IoT and Edge Computing based Manufacturing} 
	\label{fig_2}
	\vspace{-0.1in}
\end{figure*}

{As depicted in Fig. \ref{fig_2} the architecture has been divided into four domains as follows:} 
\begin{enumerate}

\item[{\bf a.}] {{\bf The application domain} is responsible for providing a comprehensive oversight over the entire manufacturing system to aid in the active administration of the system. This oversight includes services such as monitoring of data flow and network health, as well as the capacity for control of the system. The application domain, therefore, allows the system to provide flexible, generalized, and inter-operable intelligent applications while also aiding in the maintenance of service security.}

\item[{\bf b.}] {{\bf The data domain} is responsible for providing services such as data cleaning, feature extraction, and intelligent inference, which enables the system to optimize system operations so as to improve the system's throughput and efficiency. Another important feature of this domain is that it allows end nodes to quickly access data, due to the proximity of the edge computing node and the end devices, which aids in generating real-time responses for specific events. Therefore, this is a critical part of dynamically controlled manufacturing systems.}

\item[{\bf c.}] {{\bf The network domain}, in essence, is responsible for connecting the end devices with the data platform and this domain utilizes the Software Defined Networking (SDN) architecture \cite{manu2} to manage operations involved in the control plane and network transmission. A Time-Sensitive Network (TSN) protocol is also employed within this domain to handle time sensitive information and is used extensively in processing the information related to the network in sequence. This domain also offers universal standards for sustaining and supervising the time sensitive nodes, making it a critical part of the overall system architecture.}

\item[{\bf d.}] {\bf The device domain} refers to the devices located or embedded within the field apparatus like machine tools, controllers, sensors, actuators, and robots. This domain must be able to sustain an infrastructure for flexible communication models in order to maintain a variety of communication protocols by maintaining nodes which change the system's execution strategies dynamically based on the inputs obtained from the sensors. We normally observe that on the edge nodes, the information model is built with popular protocols such as OPC UA \cite{manu1} and Data Distributed Service (DDS) \cite{dds}. Finally, the unified semantics of information communication are realized within this domain of the system architecture, and it is also responsible for maintaining data privacy and security.

\end{enumerate}

\subsubsection{{Active Maintenance Case Study}}
{With the proliferation of cyber physical systems, a wide variety of industrial projects are being migrated to edge computing based frameworks because of the promise of improved efficiency, ease of maintenance, and real-time adaptability offered by this computing paradigm. We shall be reviewing a case study on a customized production line for candy packaging, as entailed in \cite{manu3}. In this study, a private cloud was used to provide service to customer orders. In order to make stable and high speed communications possible, an ad-hoc network was built connecting the edge nodes. Furthermore, in order to achieve proper exchange of information, a standardized version of the DDS protocol and ethernet were integrated before the deployment of the system. The functioning of the system can be summarized as:}
\begin{enumerate}
	\item [{\bf i.}] {Candy packaging tasks were associated with each robot and these tasks were also linked to the cloud. After getting their assigned tasks, the robots were required to pick up the particular candy assigned to them and keep the candy into the relevant open packaging. In this operation, backbone network nodes were represented by the robots.}
	\item [{\bf ii.}] {System was also capable of shifting nodes to different positions on the production line in case of any failures. Therefore, a system with multiple agents was established to improve the self-governing functionality in this scenario.}
	\item [{\bf iii.}] {The agents of the system, physically represented by robots, were independent and self-directed which means that their objective and behaviour was not constrained by other agents of the system.}
	\item [{\bf iv.}] {This system of multiple agents was deployed in order complete tasks efficiently by assigning different agents with various tasks and procedures.}
	\item [{\bf v.}] {CNP (Contract Net Protocol) was used to assign different tasks to different agents by using techniques such as winning modes, bidding and open tendering.} 
	\item [{\bf vi.}] {By the means of contests and discussions the agents are able to bargain and resolve their conflicts and so this self-organized system is able to efficiently complete the assigned tasks.}
\end{enumerate}
{The implementation of this scenario was made possible with various setups, which include the following:}
\begin{enumerate}
\item[{\bf i.}] {With the help of the Hadoop architecture, a distributed data processing system was built wherein at the local database level, real-time mining and analysis was performed with the help of Hadoop MapReduce and Hadoop Distributed File System (HDFS).}

\item [{\bf ii.}] {Information such as machine status and logs constituted the sensory data which was used to create a reasoning-based model which was loaded onto a Raspberry Pi system.}

\item [{\bf iii.}] {On the Raspberry Pi, an OPC UA server was made functional to perform pre-processing tasks on the transmission data that was acquired from different sensory devices. This data was raw in nature and hence, had to be transmitted safely and reliably which was made possible by the use of OPC UA server.}

\item [{\bf iv.}] {In order to integrate the data received from multiple sources, a semantic model was also built which reformed the data to maintain consistency, accuracy, and merit of the information. This semantic model used data fusion to provide generate features as inputs from the acquired data. Finally, this data was used as input to the reasoning-based model.}
\end{enumerate}

	

\subsubsection{{Tests performed}}
{In order to estimate the difference in performance obtained by using an edge computing based system instead of a centralized cloud computing system, a cloud-based system was also setup. This system had a centralized control server which managed the different agents of the system. In order to test the time of operation on the systems, both were tasked with completing the same orders under similar conditions of distribution of candy types. The number of candies to be packed were varied and the average time for robot operation completion was recorded for both systems. The results are summarized in the following two points:}
\begin{enumerate}
	\item [{\bf i.}] {With increase in quantity of orders, we observe that the self-organized version built on edge nodes is far more efficient and agile than the centralized system when the number of orders rises above 2000, as the operation completion time for the self-organized system becomes consistently lower that of the centralized system.}
	\item [{\bf ii.}] {With a stable production line, the speed of the backbone network in centralized version was observed to be around 16 Mb/s. However, after the deployment of the self-organized system, the backbone network speed dropped to around 5-6 Mb/s which represents a 65\% drop in speed.}

\end{enumerate}
{The results of this study suggest, that a decentralized and self-organizing system can become extremely useful in mass-production scenarios due to the reduced operation completion time. While the study shows that a decentralized system leads to reduction in transmission speeds within the backbone network, the system can still function efficiently as the reduced operation completion time outweighs the drop in the backbone network speed thereby increasing the effective system throughput.}

\subsection{Supply Chain Management}
Supply Chain Management (SCM) can be understood as a set of activities that are used to control, plan, and monitor the flow of products from their production to their distribution in the most efficient manner. While modern industries have already adopted cloud-based technologies to support their supply chains, an increasing number of these chains have begun to generate massive amounts of data from a diverse set of sensors and end devices located at different points along the supply chain. In such situations, it becomes impractical to store and process data in remote servers due to several reasons such as network bandwidth restrictions, large latency, and need for better fault tolerance. These restrictions, coupled with the proliferation of Radio Frequency Identification (RFID) technology, have given rise to edge computing based solutions for supply chain management.

Using the case study of a blackberry (fruit) supply chain as proposed in \cite{blackberry}, we shall attempt to explain how industries can augment their supply chain management systems to leverage the power of edge computing. The proposed system has a three layer architecture which is explained below:

\begin{enumerate}
    \item \textbf{Layer 0}: This layer includes the data producing end-devices (primarily RFID embedded sensors) responsible for generating relevant data such as the Electronic Product Code (EPC), temperature, internal pressure, humidity, air-quality, and other important parameters.
    
    \item \textbf{Layer 1}: This layer is primarily responsible for monitoring and control purposes which entails the generation of actuator commands, execution of the control logic, and generation of relevant alarms. With the use of active and smart edge nodes along with on-board decision support units, this layer aid administrators in improved quality monitoring as well as in the execution of real-time corrective actions.
    
    \item \textbf{Layer 2}: This layer consists of the traditional, centralized servers which can be used for long-term pattern recognition and analysis of offloaded sensor data, giving valuable insights which can be useful while optimizing production and distribution pipelines.
\end{enumerate}
As illustrated in the case study, the introduction of edge computing based technology can enable efficient monitoring and actuation in all three stages of the supply chain:
\begin{itemize}
    \item \textbf{In the field}: Edge nodes deployed at farms can aid in the real-time monitoring of blackberries. Through sensor information, the edge nodes can predict and notify farmers when the blackberries are ready for harvesting, thus improving shelf-life for the berries while also ensuring that all berries are harvested at the correct time.
    
    \item \textbf{In transit}: Edge processing nodes and sensors installed in transportation vehicles can monitor various environmental parameters of berries such as temperature, relative humidity, and light. While these systems can continuously provide updates to the system managers, they can also execute instant corrective actuation methods in response to variations in environmental parameters such as controlling the air conditioning of the vehicle, adjustment of air filters, and notifying the driver about a possible opening of the vehicle doors.
    
    \item \textbf{At the packing location}: The data from the fog nodes can be used to determine the priority of cooling of incoming crates or pallets of berries which can enhance the freshness of the products while also minimizing any wastage resulting from spoilt berries.
\end{itemize}
This case study illustrates how an edge computing based system can drastically improve the quality of monitoring for supply chains while also offering low-latency actuation techniques for system managers. Furthermore, due to the proximity of computational resources and end-devices, the amount of data transferred to the cloud servers is reduced drastically, thereby reducing the strain on the network. This leads to an improved efficiency of these supply chains and while also resulting in reduced delays associated with the networks supporting these supply chains.

\textbf{Security Issues and Challenges}: While the SCM industry can realize immense benefits by adopting edge computing based systems, the industry must also take care of certain security issues and challenges associated with these systems before successfully implementing these systems in the real world. Apart from the general issues and challenges discussed in Section V, SCM systems are susceptible to several security threats on IoT Devices, with an important example in Injection of Information Attack \cite{inj_info} where an adversary can gain control and program end-devices to distribute bogus information with the aim of disrupting the monitoring system which can result in unwanted delays or destruction of perishable resources. In a similar manner, network-based attacks like man-in-the-middle attacks, sybil attacks, or sinkhole attacks can be used by adversaries to alter transmitted data which can lead to false alarms of degraded environmental conditions or suppression of genuine alarms that can destroy the products of the supply chain. 

\subsection{Food Industry}
Modern food manufacturing industries have started to rely heavily on automated food production systems in factories to improve the quality and speed of production of consumable items. However, unlike other industries, the food industry constantly deals with perishable items - whether it is milk or sugar as raw materials or chocolates as finished products. Therefore, the food industry must invest in resources and systems that help in product traceability in all stages of production, processing, and distribution. These resources not only aid in the optimization of the manufacturing and distribution pipeline but also enable the industry to perform product recalls (such as in the case of some contamination) with minimal losses. In this regard, edge computing solutions have emerged as viable frameworks due their distributed nature and the introduction of these systems can be extremely beneficial for the food manufacturing industry.

In the system proposed in \cite{food}, food manufacturing industries can rely on QR codes, barcodes, RFID tags, or transponders implanted onto objects such as primary and secondary packaging, pallets, trucks or containers, throughout the supply chain to aid in their identification and tracking along the production and supply pipeline. Edge-computing enabled sensors can be used in the process of product identification at different points along the production and supply pipelines to ensure that the flow of products is maintained. Within this system, the edge devices can rely on ad-hoc networks to communicate with each other to determine bottle-necks along the production and supply pipelines and automatically optimize these pipelines. The centralized cloud database can also be linked with this ad-hoc network and can maintain a global database of the products for administrative supervision. Therefore, with the use of such an edge-computing powered system, the industry can rely on a latency-sensitive system that functions with reduced response times unlike a traditional cloud computing based system.

\subsection{Distributed Synchronization Services}
\begin{figure}[ht]
	\includegraphics[width=\linewidth]{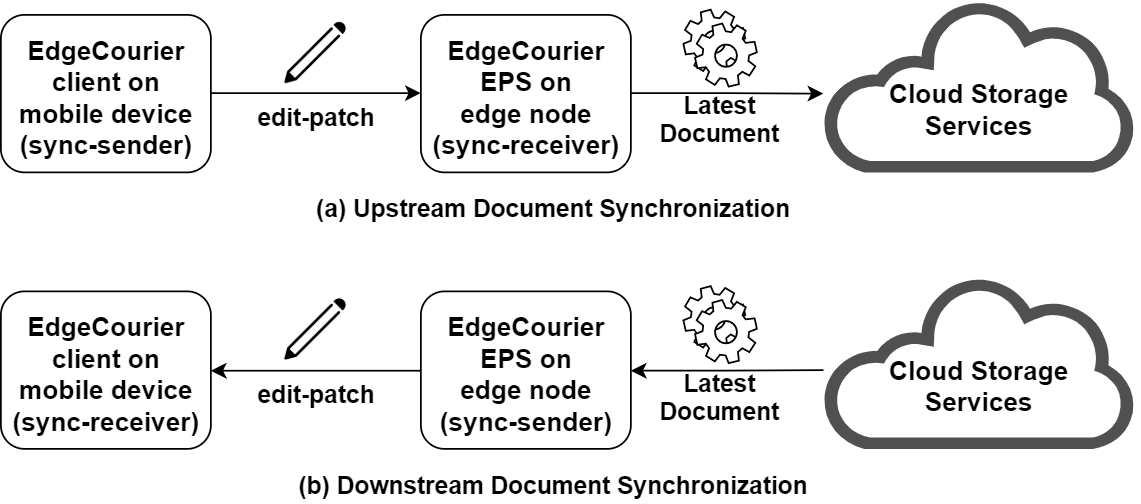}
	\caption{System overview for EdgeCourier}
	\label{fig_17}
	\vspace{-0.1in}
\end{figure}

One of the biggest use cases of cloud computing based storage is distributed data storage, commonly referred to as cloud storage services wherein files can be accessed from anywhere on the planet by connecting a system with cloud storage servers which periodically synchronize data on different devices to enable access of files. However, even for small applications like office suite softwares, cloud storage services can often lead to unnecessary bandwidth consumption while also compromising on latency. The EdgeCourier \cite{edgecourier} is file storage solution which can overcome the problems of traditional cloud computing based distributed storage options by making use of the edge-hosted personal services (EPS) technique in conjunction with the \textit{ec-sync} incremental synchronization approach. The essence of EPS is to make use of computational resources on the edge nodes (like access points or base stations) to provide localized services for mobile wireless users connected to these edge nodes. The \textit{ec-sync} synchronization approach requires two participants: the \textit{sync-sender} and \textit{sync-receiver}, both of which are instrumental in the synchronization process which is explained as follows:
\begin{itemize}
    \item The \textit{sync-sender} detects if there is any document which requires synchronization with the receiver and is responsible for capturing the changes made within the document, by going through every sub-document within the document.
    \item In order to capture sub-document changes, the \textit{sync-sender} compares two files: the edited document and the last-synced version of the same file.
    \item Thereafter, the \textit{sync-sender} places the detected changes into a file known as \textit{edit-patch}, which is transmitted to the \textit{sync-receiver}.
    \item Upon receiving the \textit{edit-patch} file, the \textit{sync-receiver} applies the edit-patch differences to the relevant sub-documents from the last-synced version of the same file to obtain the edited document.
    \item This edited document is then also shared with the cloud storage services in order to transmit it to various EPS instances or nodes across the network for global synchronization.
\end{itemize}
Furthermore, an important advantage of having different EPS instances is that they can can be managed by a centralized management service (on a cloud service), which can migrate data to and from the edge nodes if needed. This, therefore, leads to better oversight and increased fault tolerance as data can be migrated to different resources for analysis or in response to outages experienced at edge nodes. The overview of the EdgeCourier system can be seen in Fig. \ref{fig_17}.
Laboratory based studies on the Edge Courier system \cite{edgecourier} showed that with the rise in the size of documents that need to be synchronized, the time spent on network transmission becomes notably lower for the EdgeCourier system as seen with a document size of 1 MB which takes 0.6 seconds lesser on the EdgeCourier system than on the direct sync system. Such distributed synchronization systems can be particularly useful in the software development industry for real-time code synchronization in large team projects. Similarly, the banking industry can also derive some critical applications from these systems such as in the real-time synchronization of transactions and other banking data. These examples clearly show that edge computing powered data synchronization systems find a lot of applications in modern industries which require low-latency and reliable network services. As we have seen, these systems lead to reduced data transmission over the network, resulting in reduced latency and lesser strain on the network's bandwidth capabilities, hence leading to dependable network services.

\begin{figure*}[ht]
	\centering
	\includegraphics[width=130mm]{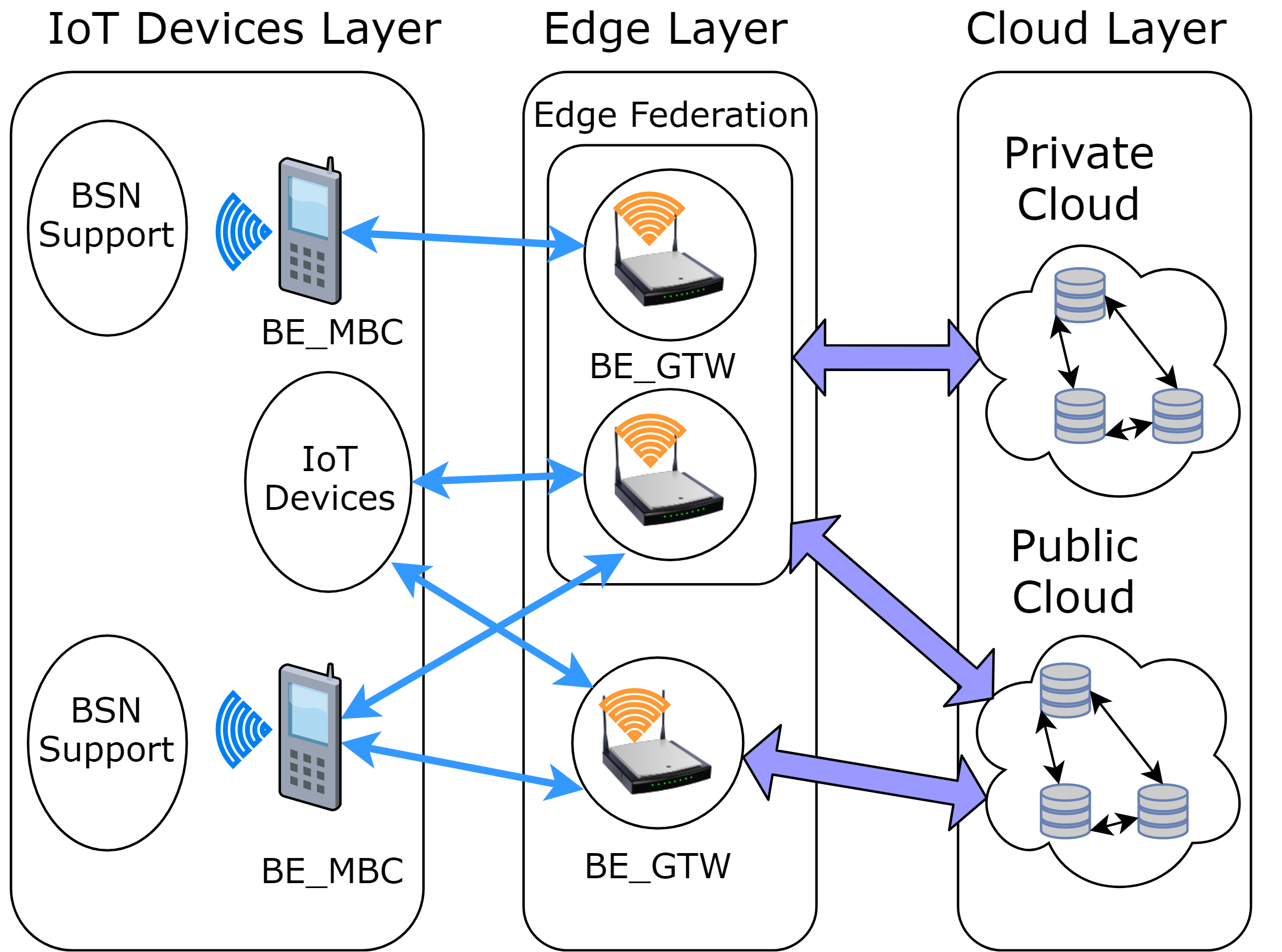}
	\caption{The BodyEdge \cite{health} Architecture} 
	\label{fig_25}
	\vspace{-0.1in}
\end{figure*}
\subsection{Healthcare}
With the recent advancements made in the domain of medical IoT devices, the healthcare industry has started to adopt IoT solutions that provide vital medical services such as the monitoring of Electrocardiogram (ECG) data and processing of Magnetic Resonance Imaging (MRI) data. However, most of the traditional IoT based solutions for healthcare rely heavily on cloud-based processing as well as storage which has started to create problems for these solutions as the massive amount of data being generated is straining the communication network’s capacity. This often leads to unpredictable delays in communication while also promoting increased latency in the network which can significantly impact healthcare operations within the hospital or clinic especially in time-sensitive situations that require urgent reactions such as in heart attacks or strokes. Therefore, modern medical IoT systems require a flexible multi-level network architecture which can cohesively work with heterogeneous sensors and process the relevant data with minimal latency to produce relevant results and responses. These requirements have led to the adoption of the edge computing paradigm in medical IoT systems due to the benefits it can provide in terms of reduced latency and improved reliability, both of which are critical for these systems. In this subsection, we will be reviewing the BodyEdge architecture \cite{health} as shown in the figure below, which is structured and inspired by the edge computing paradigm to achieve the following goals:
\begin{itemize}
    \item Reduced communication delay and latency.
    \item Wide support for scalability and responsiveness.
    \item Limited cost in terms of bandwidth for data transmission (i.e. only limited statistics data needs to be transmitted to the cloud).
    \item Improved Privacy (since the edge network may be interpreted as a private cloud).
\end{itemize}
This architecture consists of two complementary parts. The first, is a mobile client called BodyEdge Mobile BodyClient (BE-MBC) which is primarily responsible as a relay node for communication between the sensors and the edge client using multi-radio communication technology. The second is a performing gateway known as the BodyEdge Gateway (BE-GTW), which is placed at the edge of the network and is primarily responsible for acquiring device data and locally processing it to produce valuable insights and patterns that can be relayed back to the end devices or sensors. In addition to this, the gateway also ensures communication with the cloud to allow users to maintain oversight over this system.

In order to validate the BodyEdge architecture, it was physically implemented in \cite{health} and compared with a cloud based architecture for the task of stress detection using cardiac sensors. Within the implementation, the BE-MBC module was installed on a smartwatch which was paired with a chest band to acquire ECG signals. The BE-GTW was installed on an independent hardware platform (Raspberry Pi3) as well as on an Azure cloud virtual machine in order to perform the comparative study. Finally, the edge-based system with the BE-GTW installed on the Raspberry Pi3 was tested on 100 athletes to determine stress levels using the Heart Rate Variability (HRV) technique \cite{hrv} and the average round trip delay time (RTT) for this case was 152 ms. The same experiment was then conducted with the cloud-based system which yielded an average round trip delay time (RTT) of 338 ms. This result, therefore, corroborates our assumptions about the performance benefits offered by edge-computing based systems in terms of reduced latency and indicates that medical IoT systems should indeed adopt edge computing based network architectures.

\subsection{Agriculture}
\begin{figure}[ht]
	\includegraphics[width=\linewidth]{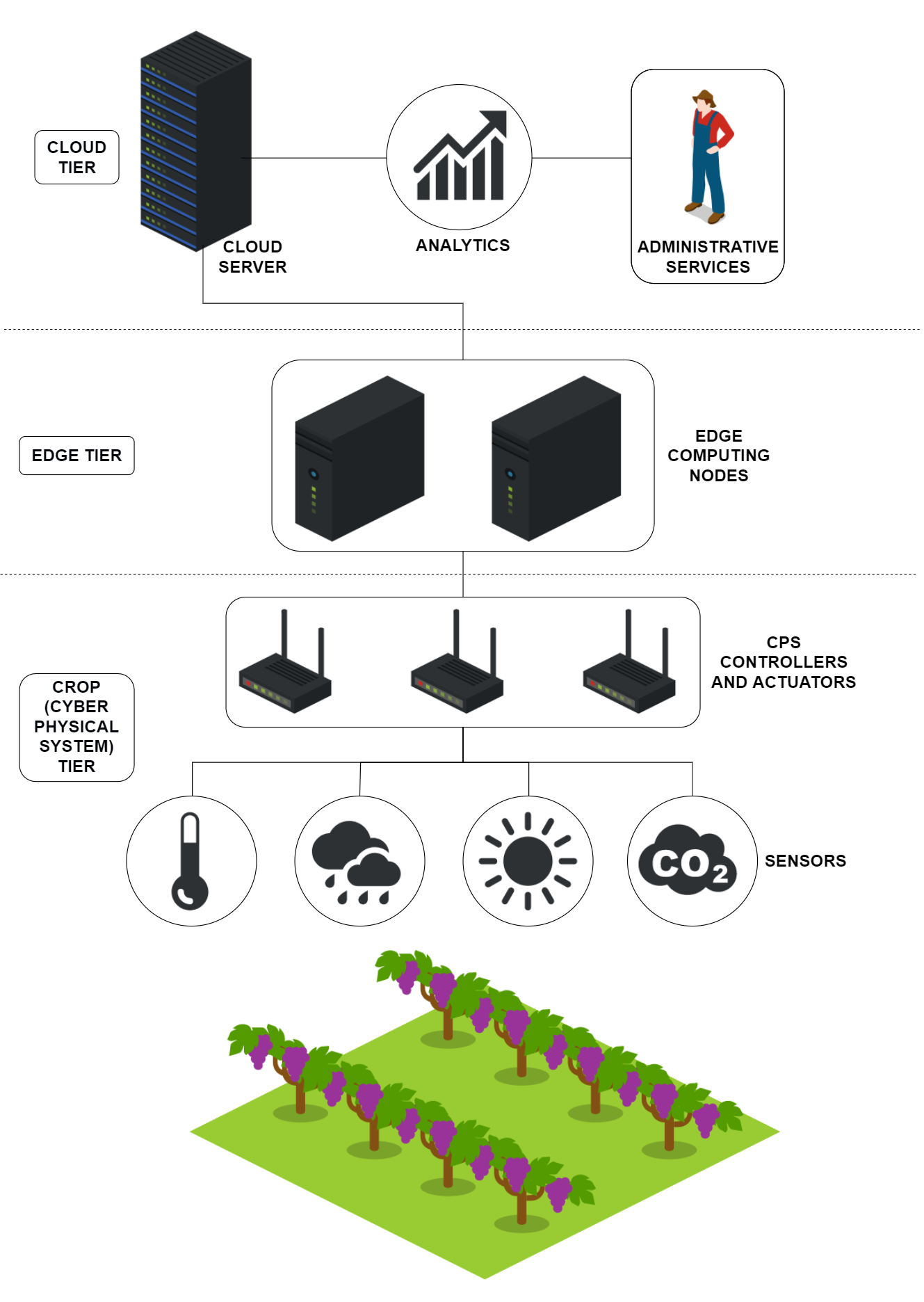}
	\caption{Architecture Overview for Agricultural Monitoring System}
	\label{fig_18}
	\vspace{-0.1in}
\end{figure}

Modern agriculture has extensively embraced automation and modern technology so as to improve and optimize existing agricultural processes due to the improved connectivity between agricultural resources. As technology is becoming increasingly interconnected, edge computing based infrastructures have started to dominate most network-based applications and in order to tackle the growing amount of data being generated by end devices, the agricultural industry has also started edge computing based architectures in order to create latency-sensitive applications for agricultural processes. The concept of Precision Agriculture (PA) has seen a significant rise in popularity due to the improvement in sensor technologies, and several systems based on edge computing have been proposed, like the precision agriculture platform \cite{agro}. These systems make use of intelligent algorithms in conjunction with smart sensors and actuators in the field to providing real-time monitoring services that enable control services to maintain optimal environments for crop growth. In the system proposed in \cite{agro}, the architecture is divided into 3 tiers namely: crop (Cyber Physical System or CPS) tier, edge computing tier, and the cloud tier. The architecture has been illustrated in Fig \ref{fig_18}.
The crop (CPS) tier is majorly comprised of sensors that aid in real-time monitoring of various environmental parameters such as temperature, humidity, pH, $CO_{2}$ levels, solar radiation, and other important factors. In addition to sensors, this tier also supports various actuation devices such as soil nutrition pumps, valves, irrigation devices, ventilation devices, and light-control devices. Within this architecture, operations at this tier require low latency and high reliability in communication so that emergency services can be enacted without human intervention, which is made possible through the edge computing based computational nodes situated closer to the data sources. In continuation, edge nodes within the edge computing tier are responsible for executing commands through actuation devices based on inputs received from sensor networks in the crop tier. Therefore, this layer is responsible for control of irrigation, climate control, nutrition control, and other auxiliary tasks like alarm and energy management. Finally, the cloud tier is responsible for long-term data analytics and system management services. The physical implementation of this system showed savings of more than 30\% in terms of water consumption along with savings of nearly 80\% in terms of some soil nutrients when compared with a regular open crop. In addition to environment monitoring, edge computing powered systems can also be employed for video analytics through UAVs that can help farmers in optimized weeding and harvesting. This clearly illustrates the impact of automation on the agricultural industry, and shows how edge computing based architectures can replace cloud computing frameworks especially in applications that require low-latency and high reliability.

\section{Industrial Applications of Fog Computing}
\subsection{Smart Grids}
Conventional energy grid systems have been powering industries and countries for the past 100 years, and with the tremendous rise in demand for electrical power, the domain of IoT has emerged to be the pioneering technology that is leading developments in the smart grid systems. Traditional grid operations relied on simple analog meters to record units of power flowing per month to each household or industry, but with the evolution of intelligent and autonomous systems, modern smart grids offer solutions that allow comprehensive oversight over energy distribution which is beneficial to both consumers and producers. With respect to power producers, these smart grid solutions allow accurate monitoring of energy demands and supplies which allows them to effectively control pricing as well as load balancing to sustain the healthy functioning of the grid. On a similar note, consumers can monitor their energy consumption in real-time with respect to each device which allows them to effectively and reliably manage their energy spending. The framework of such a smart grid, therefore, involves a heavy dependence on the collection and aggregation of real-time data from every device within each household or industry that is powered by the grid. This will inevitably lead to the generation of a large amount of data that needs to be efficiently managed and analyzed while maintaining the security of the data. In order to manage such massive amounts of data, it is easy to perceive that the cloud computing paradigm cannot be a viable network architecture for these IoT powered smart grid solutions since the sheer volume of the data would not adhere to any conventional network’s transmission capacity. In order to reduce the strain on the network capacity, fog computing-based grid systems can become a viable option since the fog computing architecture allows computational offloading from the centralized cloud servers to fog nodes that are situated closer to the end devices. This distributed nature allows the network to function with low latency and improved reliability while also maintaining data security, and these are exactly the properties that a modern smart grid system requires.

\begin{figure}[ht]
	\includegraphics[width=\linewidth]{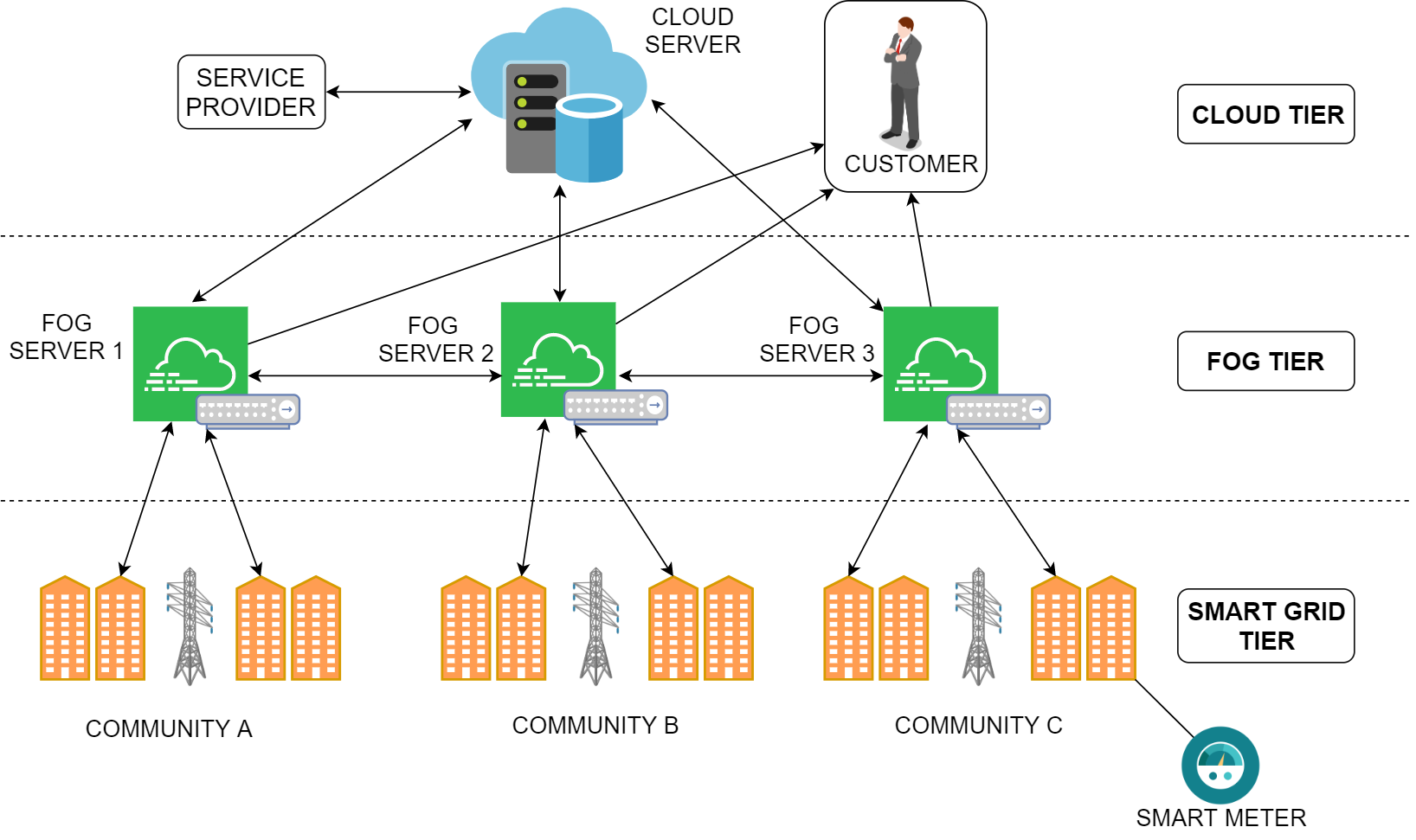}
	\caption{Structure of Fog Computing enabled Smart Grid} 
	\label{fig_20}
	\vspace{-0.1in}
\end{figure}

The basic architecture of smart grid systems is generally composed of advanced metering infrastructure (AMI) along with area networks, data centers, and integrated substation centers. Within this architecture, AMI ensures two-way communication is maintained between the end devices and the fog nodes which leads to a secure, reliable, and cost-effective service. The model proposed in \cite{grid} is a three-tier architecture as shown in Figure \ref{fig_20}.

The first tier is comprised of the smart meters which are responsible for collecting data regarding energy consumption as well as for inter-tier and intra-tier communication. The second tier comprises the resource-rich fog nodes which are responsible for delivering the majority of computational services to the network. Finally, the third tier comprises the traditional cloud servers which are usually responsible for oversight and maintenance of the entire grid. This structure allows inter-tier communication within the first and second tiers which enables different geographical sub-grids to communicate with one another. 

Through the following points, we can appreciate the benefits offered by fog computing architecture: 
\begin{itemize}
    \item The smart metering technology enables the energy producers to monitor power loads in real-time which helps them in drafting an effective load-balancing methodology, with extremely low latencies and transmission delays.
    
    \item The smart meters allow consumers to monitor the energy consumed by each device in real-time and this can aid them in controlling device usage dynamically to minimize their energy costs.
    
    \item While the smart meters maintain a local database of the profile of energy consumed by each individual device, they actually aggregate this data for the complete household or industry and forward this encrypted aggregate to the fog servers. These fog servers can then store this data securely within storage systems that are localized in that geographical area, and because the encryption key is only known to the fog node and the respective smart meter, the system maintains privacy even if the data is accessed by someone through the cloud server.
    
    \item Finally, the varied geographical location of fog computing nodes can be beneficial to the grid in an interesting way: specifically for the case of electric vehicles which can be charged at any location inside the grid while the grid maintains the correct billing information. For instance, if an electric vehicle is charged in any neighbourhood, the smart meter deployed in that neighbourhood can identify the owner of the car using a unique identifier and transfer the billing information via the fog node tier to the owner’s smart meter, thereby ensuring consistency in billing within the smart grid.
\end{itemize}

\subsection{Communication}
\begin{figure}[ht]
	\includegraphics[width=\linewidth]{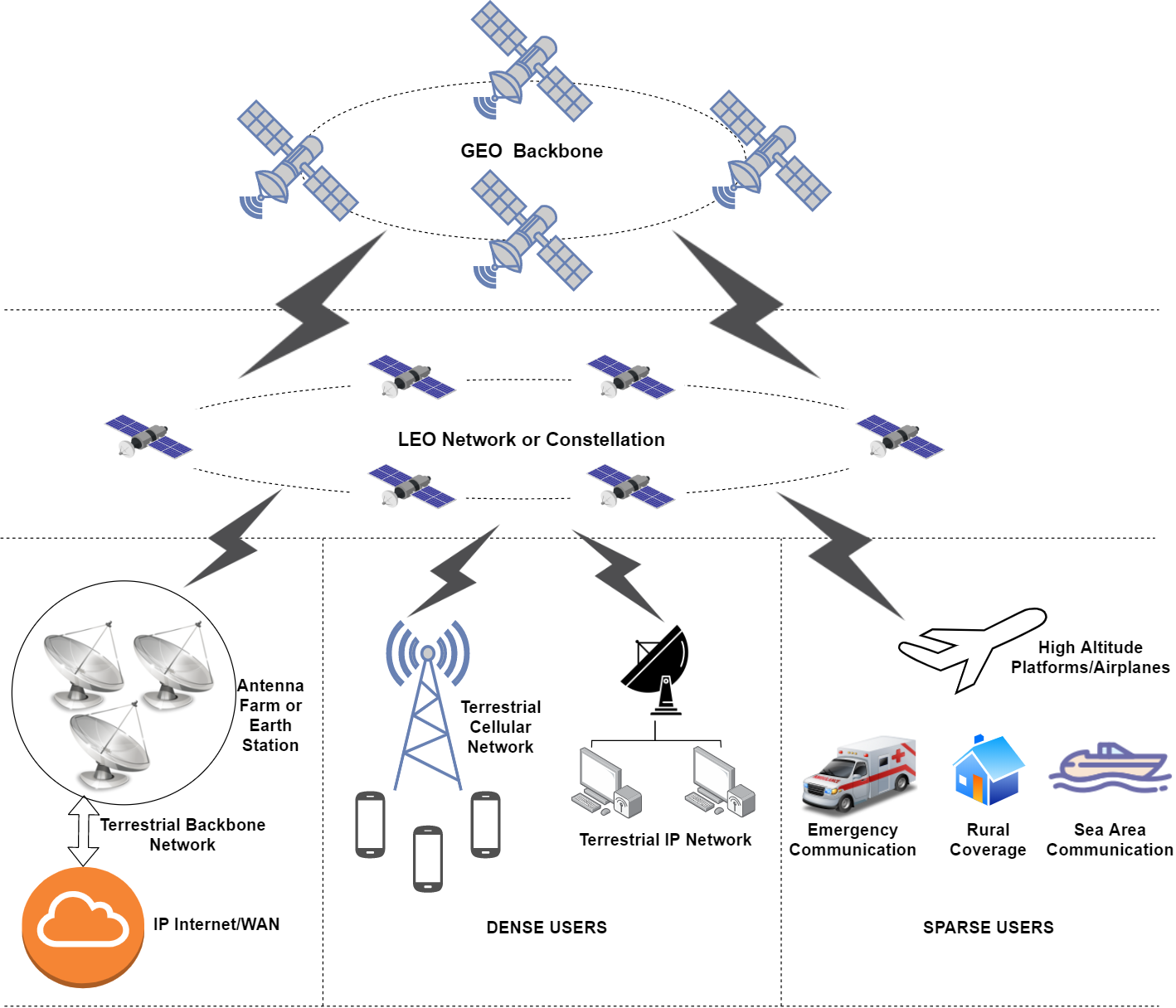}
	\caption{Traditional Satellite Terrestrial Network}
	\label{fig_28}
	\vspace{-0.1in}
\end{figure}

\begin{figure}[ht]
	\includegraphics[width=\linewidth]{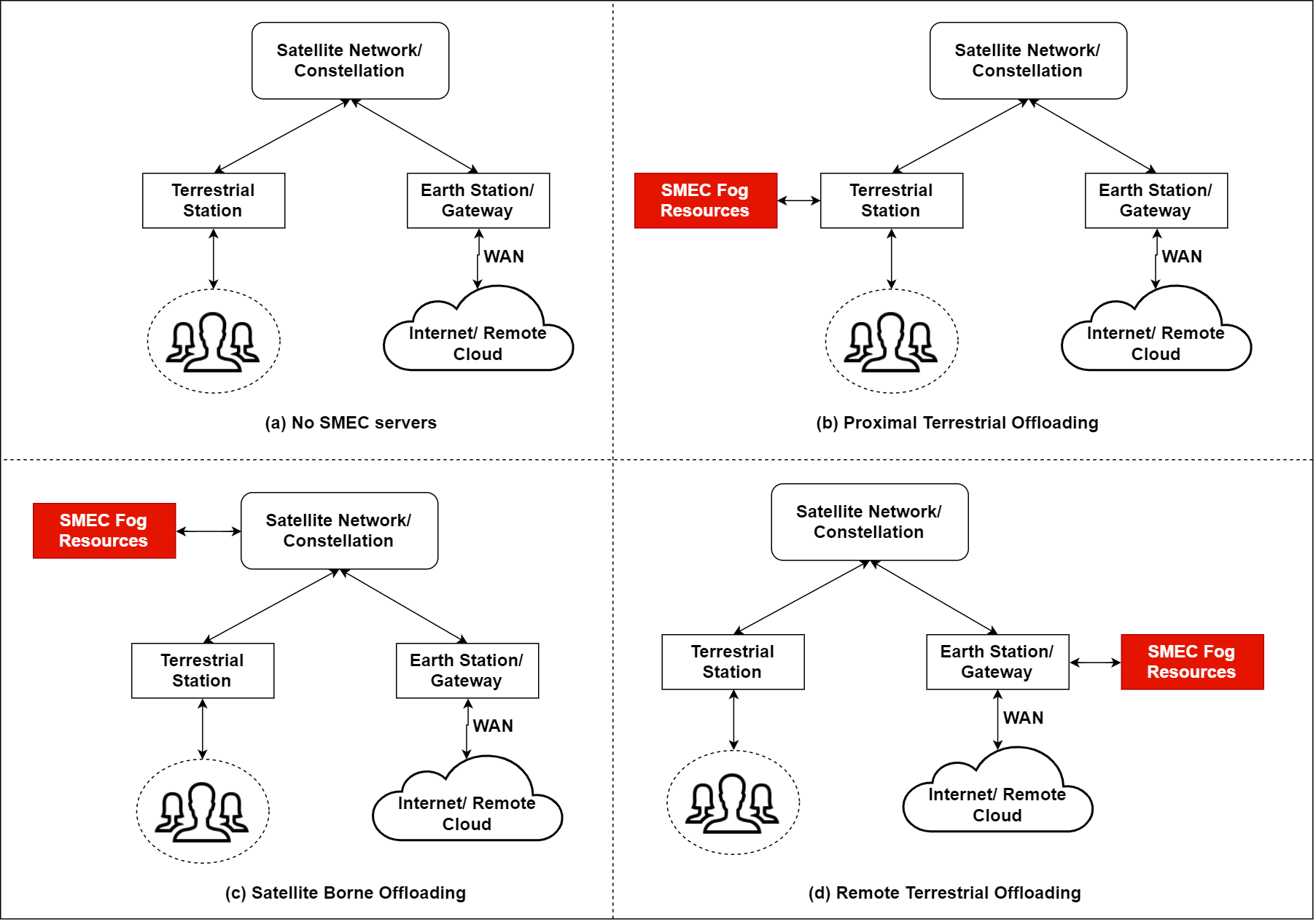}
	\caption{SMEC with offloading at different fog sites}
	\label{fig_26}
	\vspace{-0.1in}
\end{figure}

With the recent advances made in satellite technology, the communication industry has started to rely heavily on satellites to provide access to people situated in remote locations. Satellite-Terrestrial Networks (STN) are communication networks that have emerged as one of the most promising low-cost technology which can lead to ubiquitous access to internet connectivity across the globe. A majority of STN setups make use of Low Earth Orbit (LEO) satellites to provide connectivity to sparsely distributed users by interconnecting small terrestrial terminal stations which are placed in remote locations to ensure maximum area coverage, as shown in Figure \ref{fig_28}.
But, with the evolution of smartphones and tablets, the amount of data that needs to be transferred across the network has increased drastically, particularly because of an increase in the number of applications such as speech recognition and gaming that make use of cloud services to process user-generated data. This puts a strain on the network’s data transfer capacity, and so we must look towards computational offloading to help alleviate this problem. In this situation, satellite mobile edge computing (SMEC) \cite{smec} can be a possible solution that can offload computation as well as storage to local servers, thereby leading to an improved QoS, increased reliability, and reduced latency. This technology, although dubbed as edge computing, is better classified as a fog computing based technology as computational resources are essentially an extension of the cloud servers. Therefore, the introduction of fog computing resources near the end-devices can lead to content caching and other storage facilities which effectively reduces that traffic in the overall STN. In terms of computational offloading, the fog sites can be located at 3 different locations, and these are:
\begin{itemize}
    \item \textbf{Proximal Terrestrial Offloading (PTO)}: In this situation, satellite mobile fog computing servers are deployed at terrestrial stations, as shown in Figure \ref{fig_26} (b). The advantage of this system is that the communication latency is significantly reduced because backhaul transmission through the satellite is avoided. While such a system would be extremely useful for terrestrial terminal stations that cater to dense user areas, it would not be practical for terrestrial terminal stations that are placed in spare user areas especially because these stations do not hold extra computational facilities and are remote.
    
    \item \textbf{Satellite Borne Offloading (SBO)}: In this situation, the satellite mobile fog computing servers are deployed in LEO satellites, as shown in Figure \ref{fig_26} (c). With this network extension, both sparse and dense users will benefit from reduced latencies while the traffic in the terrestrial backbone network will also reduce significantly. However, the latency in this situation would be higher than that of PTO and it would significantly increase the power consumption of satellites which will be performing the offloaded computations which will not be practical for satellites with limited power sources.
    
    \item \textbf{Remote Terrestrial Offloading (RTO)}: In this situation, the satellite mobile fog computing servers are deployed to the terrestrial backbone network, as shown in Figure \ref{fig_26} (d). In this situation, the delays in transmission over the WAN IP that connects with the Remote Cloud servers can be avoided and this translates to a reduced latency when compared to the situation with no edge computing offloading. The latency in this network scheme is higher than PTO and SBO, but it is the most practical scheme to implement and maintain.
\end{itemize}

\subsection{Manufacturing Process Monitoring}
With rapid globalization, industries across the globe have started to adopt modern process control systems which rely heavily on sensor networks that efficiently monitor production lines and processes while collecting valuable data which can be used to identify faults before they occur while also aiding in optimization efforts so as to improve the throughput and performance of the industry. In this regard, we shall be looking at fog computing-based framework for process monitoring in different production environments. The proposed system architecture in \cite{process} is described in a sequential manner:
\begin{itemize}
    \item Step 1: Collect machine data from the production environment that streams real-time data from various sensor networks and communication adapters that function on protocols such as Simple Object Access Protocol (SOAP), MTConnect, and Open Platform Communications Unified Architecture (OPC UA).
    
    \item Step 2: Stream the raw data to a private computational fog node which is responsible for real-time monitoring and providing time-sensitive control signals to the production environment. This allows the system to function with low response times, improves reliability, and reduces the strain on the network's capacity as data is processed in a fog computing node that is situated close to the production environment.
    
    \item Step 3: In addition, various samples from this data can be sent to high-performance cloud data centers which can be used to build models for predictive maintenance and process optimization. Since these samples are small in size and sporadically transferred to the cloud, the strain on the network's capacity is minimal while the models built with the sampled data can be extremely beneficial for the industry in terms of improved throughput and reduced unplanned downtimes.
    
    \item Step 4: Apply these predictive models to raw data and obtain tangible insights into the production environment's real-time health and performance.
\end{itemize}
\begin{figure}[ht]
	\includegraphics[width=\linewidth]{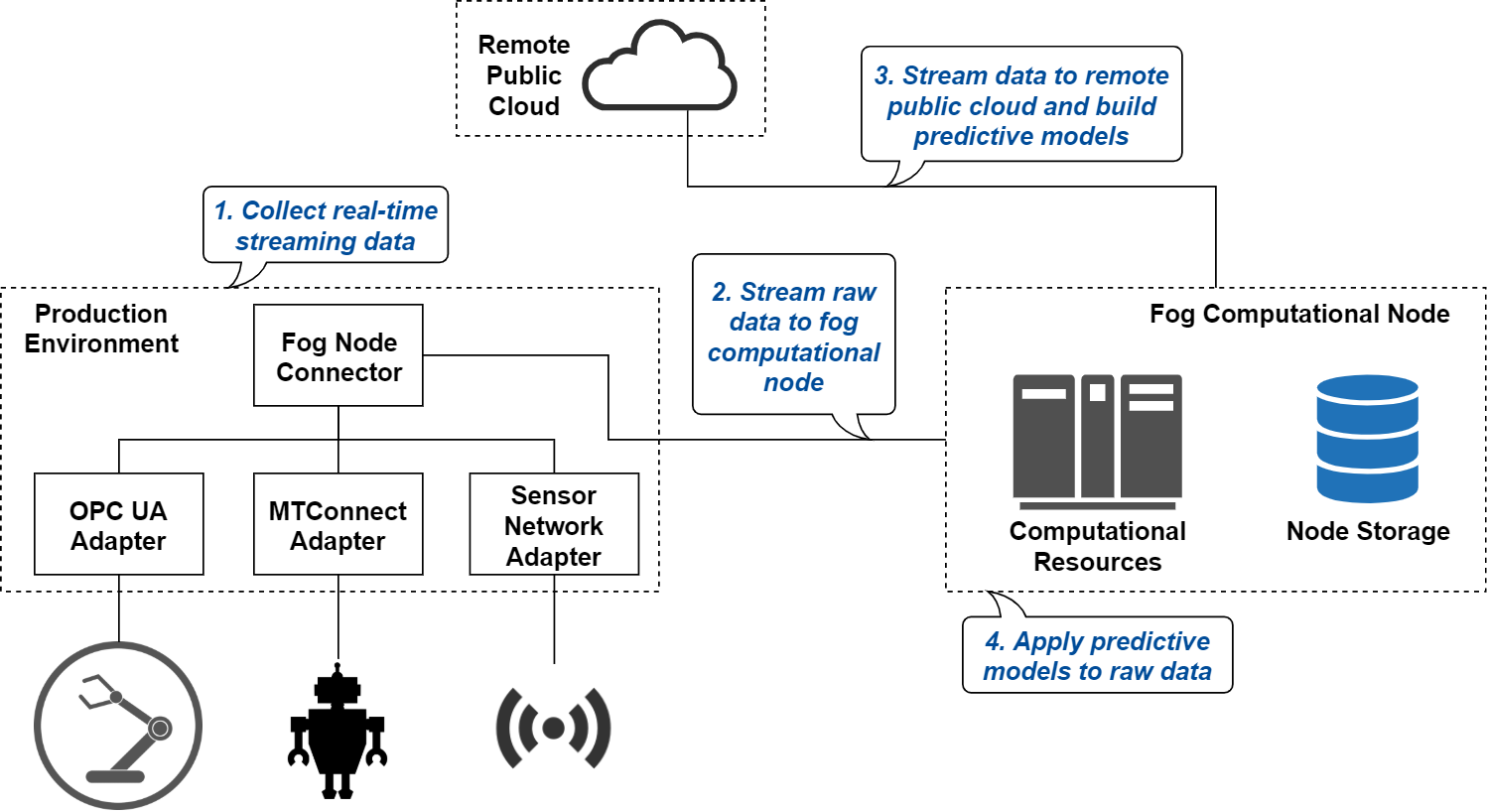}
	\caption{Architecture for the Process Monitoring System} 
	\label{fig_23}
	\vspace{-0.1in}
\end{figure}

\section{Future Directions}
The edge and fog computing paradigms are considered as powerful extensions to the cloud computing paradigm, however, they face some common challenges \cite{edgeovercloud} that are yet to be addressed. In this section, we describe the some of the major issues faced by these paradigms which can also serve as potential research directions.
\subsection{Programmability and Task Partitioning}
In the traditional cloud computing based architectures, users generally program their back-end applications on an abstract platform, without worrying about the exact configuration of the cloud server. The benefit of this abstraction is programmability, since the user is not aware about the exact configuration of the platform which means that the cloud service providers can easily compile the application and run it on a single runtime of the cloud server which can have a variable configuration, unknown to the user. However, with the rise of the edge and fog computing paradigms, back-end processing is distributed across an array of distributed computational nodes - all of which can have slightly different run-times. This creates an interesting and challenging problems for system designers, who need to design optimized methods for distributing computation as well as storage across nodes, while making sure that synchronization processes do not impact the network's transmission capacities and ensure low latency in intra-network transmissions.

An important issue that arises with the evolution of distributed computing paradigms like edge and fog computing, is the issue of task partitioning. Within these paradigms, it is imperative that the system design takes into account the optimization of task partitioning and process scheduling, to facilitate concurrent execution across distributed nodes. An optimized task partitioning scheme allows the system to autonomously locate edge or fog nodes in real-time, and allocate distributed tasks to these nodes intelligently, while taking into account various factors such as the computational power associated with each node as well as the associated network latency involved in exchanging data with these nodes.

In consideration of these issues, system designers should also think about control - whether the system should allow users to implicitly or explicitly control computational resources. In case of implicit control, which can be seen in the case of Amazon's Lambda@Edge \cite{lambda}, where the users need not worry about server administration, as the web services are responsible for running and scaling the application at resources available closest to the end users. This leads to reduced complexity of programming for the users, while giving system administrators greater control over the network. In contrast, explicit control of the network gives users greater flexibility in terms of resource allocation, which can often lead to improved efficiency and increased reliability. This explicit control, however comes at the cost of increased complexity in terms of programmability, and goes to illustrate how designers need to make trade-offs while planning the layout of edge and fog computing based systems.

\subsection{Security and Privacy}
With an increased interest in the edge and fog computing paradigms, people have started to appreciate the capabilities of these paradigms which enable the extension of storage, networking, and processing resources of cloud computing servers toward the edge of network. However, with this rise in flexibility and distribution leads to several security and privacy concerns \cite{chse} that must be addressed by system designers. After analyzing several different aspects of the network security, we can summarize the major security and privacy concerns as follows:
\begin{enumerate}
    \item \textbf{Trust and Authentication}: Edge and Fog Computing based networks are expected to provide secure and reliable services to all users and this leads to an important requirement in that all devices on the network should be able to trust one another. Therefore, trust plays a two-way role within edge and fog computing based networks. This implies that fog or edge nodes that offer services to the network must be in a position to validate whether the resources requesting these services are indeed genuine. Similarly, edge or fog nodes that are transmitting data to or requesting services from network resources should also be able to verify whether these resources are genuine or not. These concerns have given rise to various authentication mechanisms which can be used to authenticate network resources before transmissions and requests. Systems can employ mechanisms such as permissioned blockchain networks like TrustChain \cite{trust} for authentication, cryptographic authentication schemes like SAKA-FC \cite{saka}, and hardware-based authentication schemes like Physically Unclonable Functions (PUF) \cite{puf}, to authenticate network resources.
    \item \textbf{Integrity}: Edge and Fog Computing systems should always ensure that data transmission within the network should be done in a secure manner so that transmitted data is not altered or modified by attackers. The most prominent method to ensure integrity of data in networks is through the cryptographic signature verification systems like the GNU Privacy Guard (GPG) system \cite{gnugpg} which is used to digitally sign transmitted data. The received data is then verified at the receiving station to establish integrity of the data, which is extremely important in edge and fog computing based systems as they rely heavily on intra-network data transfers due to their distributed topology.
    \item \textbf{Availability}: The availability of information refers to the ability of the system to ensure that authorized parties are able to access relevant information whenever needed. The biggest concern with respect to availability of information is Denial of Service (DoS) attacks that hamper or eliminate accessibility to information. Edge and Fog Computing based systems are generally well equipped to handle DoS attacks since these systems have distributed computational resources, however Distributed Denial of Service (DDoS) attacks can still impact these systems and in order to protect networks or applications against DoS attacks, designers often make use of Web Application Firewalls (WAF), smart DNS resolution services, and other intelligent traffic management techniques to ensure service security.
    \item \textbf{Confidentiality}: The confidentiality of information represents the ability of the system to protect information from being disclosed to unauthorized parties. This implies that edge and fog computing paradigms should ensure that information is stored securely in order to prevent data leaks, which is especially likely due to the distributed architecture of these paradigms. Edge and Fog Computing based architectures often use homomorphic encryption schemes as well as cryptographic hashing techniques to store confidential data at different distributed locations within the network. Due to the use of these techniques, even if attackers are able to gain access to secure databases, they will not be able to understand the data as it will be in an encrypted format.
    \item \textbf{Data Ownership}: This issue extends from the fact that unlike cloud computing based systems, edge and fog computing based systems store data in distributed locations across the network which means that the system can store data locally at the computational nodes, thereby providing complete access and ownership to the end users. However, these paradigms often involve transmission of data between nodes especially when processing or computations have been offloaded to different nodes on the network, and this creates a problem in the data ownership. Thus system designers should take this behaviour into account while drafting the privacy policy of the network. This also involves thinking about legal jurisdictions, such as when data crosses international borders, it may be subject to different regulations. This means that data transfer methods should consider the compatibility of data with two different data regulation policies with respect to the source and destination.
\end{enumerate}
\subsection{System Metrics}
While there exist a large variety of advantages that arise due to the architecture of the edge and fog computing paradigms, there are some associated metrics which also need to be considered while designing these paradigms. Importantly, system designers often deal with the design of policies which govern task partitioning and work offloading from one computational node to others, and in such situations, they should give importance to the following metrics:
\begin{enumerate}
    \item \textbf{Energy}: Edge and fog nodes often consist of embedded devices such as wireless access points, routers, or switches, which often have power sources in the form of batteries. Due to the limited capacity of the batteries, system designers should always consider if it would be energy efficient to offload some task to a particular node, while also taking into account the computational power associated with that node and the expected amount of computation that is required for the task being offloaded. An important environmental benefit in this regard is that the energy requirement of fog and edge nodes are smaller than that of cloud servers. This means that the edge and fog nodes can use renewable energy sources for their power requirements, leading to an overall reduction in $CO_2$ emissions, which shows that the edge and fog computing paradigms are also much more eco-friendly when compared to cloud computing.
    \item \textbf{Cost}: While migrating applications to edge and fog computing based architectures often leads to reduced latency, improved reliability, and increased fault tolerance, it still comes at the expense of increased cost. With thousands of embedded computational nodes in modern edge and fog computing based systems, the cost is generally much higher than traditional cloud services, which means that systems within the edge and fog computing paradigms should be cost efficient, so as to justify their development in response to improved user experience.
    \item \textbf{Bandwidth}: The edge and fog computing paradigms need to be designed while taking bandwidth into consideration, especially in the situation of low-cost systems which generally have low bandwidths within the network. In the edge and fog computing paradigms, we see a lower amount of data transmission whenever a larger amount of data is processed closer to the edge, since no data needs to reach the remote cloud server. However, the distributed nature of the system can often increase the amount of transmissions within the system, especially in co-operative systems that rely heavily on inter-node communication. Therefore, system designers can appreciate the two major factors that influence bandwidth consumption and can organize their system accordingly.
\end{enumerate}

\section{Summary and Conclusions}
With the recent advances within the domain of IIoT, people have started to observe strong trends which indicate a rapid growth in the number of smart devices connected to IIoT networks and this growth cannot be supported by traditional cloud computing platforms. In response to this, edge and fog computing systems have emerged as important frameworks which have the potential to support the growing demands of automation in different industrial settings. As these paradigms are inherently distributed in nature, their resources are distributed along the edges of the network which leads to reduced latency and improved reliability of services associated with systems designed using these paradigms. Through this paper, we have described the fundamentals of the edge and fog computing paradigms while comprehensively exploring the benefits offered by these systems over traditional cloud-based platforms. Furthermore, the paper details several industrial applications for both edge and fog computing through an in-depth analysis of proposed system architectures for the different industrial use cases. With several supporting case studies and experiments explained in the paper, we practically demonstrate the superiority of these computing paradigms and build a strong case for the adoption of these paradigms in modern industrial systems. Finally, we present the major issues and challenges faced by these paradigms, along with some plausible solutions which serve as future research directions.


\end{document}